\documentclass[]{article}

\pdfoutput=1
\usepackage[latin1]{inputenc} 
\usepackage{cite}

\usepackage{graphicx}
\usepackage{mathrsfs}
\usepackage{amsmath,amssymb}
\usepackage{hhline}
\usepackage{dcolumn}
\usepackage{bm}

\usepackage{color}

\title{Random walk centrality for temporal networks}
\author{Luis E. C. Rocha\dag~and Naoki Masuda\ddag \\
\small \dag~Department of Public Health Sciences, Karolinska Institutet, Solna, Sweden \\
\small \dag~Department of Mathematics, Universit\'e de Namur, Namur, Belgium \\
\small \ddag~Department of Mathematical Informatics, The University of Tokyo, Tokyo, Japan \\
\small \ddag~CREST, JST, Saitama, Japan \\
\small \texttt{masuda@mist.i.u-tokyo.ac.jp} }
\date{\today}

\begin{document}
\maketitle

\section*{Abstract}

Nodes can be ranked according to their relative importance within the network. Ranking algorithms based on random walks are particularly useful because they connect topological and diffusive properties of the network. Previous methods based on random walks, as for example the PageRank, have focused on static structures. However, several realistic networks are indeed dynamic, meaning that their structure changes in time. In this paper, we propose a centrality measure for temporal networks based on random walks which we call TempoRank. While in a static network, the stationary density of the random walk is proportional to the degree or the strength of a node, we find that in temporal networks, the stationary density is proportional to the in-strength of the so-called effective network. The stationary density also depends on the sojourn probability $q$ which regulates the tendency of the walker to stay in the node. We apply our method to human interaction networks and show that although it is important for a node to be connected to another node with many random walkers at the right moment (one of the principles of the PageRank), this effect is negligible in practice when the time order of link activation is included.

\newpage

\section{Introduction}

Random walks of various types are prototypical dynamical processes on networks. Random walk models are not only objects of pure theoretical interest but the study of their dynamics enlightens general properties of diffusive processes. For instance, properties of the random walks are tightly connected to those of interacting particle systems such as stochastic opinion formation models~\cite{Liggett1985book,Durrett1988book} and to current flow on electric circuits~\cite{Doyle1984book}. Furthermore, random walks have been applied to searching and routing on networks~\cite{Kleinberg2000Nature,Adamic2001PhysRevE,Guimera2002PhysRevLett,Franceschetti2006JApplProb,Draief2006JApplProb}, detection of network communities~\cite{Rosvall2008PNAS}, and respondent-driven sampling~\cite{Salganik2004SocMet,Volz2008JOffStat}. A particularly successful application is on ranking of nodes. The PageRank algorithm used for ranking websites and other entities is equivalent to the stationary density of a random walk~\cite{Brin1998conf,Langville2006book}. Other definitions of centrality (i.e.\ ranking) of nodes in networks on the basis of the random walk have also been proposed~\cite{Noh2004PhysRevLett,Newman2005SocNetw,Callaghan2007AmerMathMonthly,Saavedra2010PhysicaA}.

Previous research mostly focused on static structures, i.e.\ snapshots of networks where the links between the nodes are fixed. Nevertheless, various networks in which node ranking is relevant are dynamic, meaning that a link is used only occasionally in time. The structure of the web graph, for instance, is continuously fluctuating with webpages and links being added and removed at every moment~\cite{Desikan04}. Human interaction networks derived from, for example, email communication, face-to-face conversations, and  sexual contacts, are highly dynamic and follow irregular temporal patterns. As a consequence, the respective interaction matrices vary over time, and a static network representation of such systems becomes deficient. Such varying structures, in which the time order of link availability is relevant, are collectively called temporal networks~\cite{HolmeSaramaki2012PhysRep} in contrast to aggregate (or weighted static) networks, in which all interactions within a time-window are collapsed into weighted links.

In the present paper, we propose a centrality measure, named TempoRank, for temporal networks on the basis of the random walk. To realize that, we have to formulate and characterize random walks on temporal networks. Previous studies have addressed diverse properties of the dynamics of random walks on temporal networks, for instance, the cover time~\cite{Avin2008LNCS}, mean first-passage time~\cite{Acer2010MobiOpp,Perra2012PhysRevLett}, the stationary density~\cite{Perra2012PhysRevLett}, mixing time~\cite{Delvenne2013arxiv}, conditions for stationarity and ergodicity~\cite{Figueiredo2012SIGMETRICS}, and properties of the so-called active random walk~\cite{Hoffmann2013chapter,Hoffmann2012PhysRevE}.

However, to apply a random walk centrality measure to real data, we have to understand random walks on real temporal network data. This is nontrivial for at least two reasons. First, available data are ubiquitously non-stationary. Second, with a high temporal resolution, a snapshot of a network at each time is often sparse, which limits possible pathways for random walkers such that the walk is no longer completely random. In this data-driven direction, Starnini and colleagues analyzed the coverage and the mean first-passage time of a random walk model on temporal network data. They found that the diffusion was slower on the temporal network in comparison to the aggregate version~\cite{Starnini2012PhysRevE}. In contrast to their work, we are interested in the stationary density of the random walk in the present study. In another study, Ribeiro and colleagues connected temporal network data to the stationary density of the random walk~\cite{Ribeiro2013SciRep}. They obtained the degree (or weighted degree, also called the strength) of the aggregate network from the data to determine the Poissonian node activity of an evolving network model. Temporal and structural patterns beyond those contained in the node degree of the aggregate network, such as the global structure of the aggregate networks and distributions and correlation of interevent times were ignored. In contrast to Ref.~\cite{Ribeiro2013SciRep}, we use temporal network data to directly define the pathways for random walkers, as done in Ref.~\cite{Starnini2012PhysRevE}.

We formulate the random walk under periodic boundary conditions and regard temporal network data as sequences of snapshots, each of which is an observation of a network within a given time window. We examine the stationary density of this random walk and argue that the local inflow considered in the so-called effective network, derived from the original network, is sufficient for accurately approximating the stationary density, or the centrality, of the nodes in the temporal networks.

\section{TempoRank: A temporal random walk centrality}

In this section, we define the TempoRank, i.e.\ the temporal random walk centrality of a node in temporal networks. TempoRank is the stationary density of the random walk under the periodic boundary condition in time. We also discuss the mixing properties, i.e.\ the conditions of convergence to a unique stationary density of the random walk.

\subsection{Temporal networks}

A temporal network with $N$ nodes and length $T$ is defined as a sequence of $r$ time snapshots of equal size $T_{\rm w}=T/r$. A temporal network data set typically consists of a list of contacts, and a contact is defined by the identities of the two interacting nodes $(i,j)$, the beginning time $t$ of the contact, and sometimes the duration $\Delta t$ of the contact. The number of contacts between nodes $i$ and $j$ that occur in the $t$-th snapshot, i.e.\ between time $(t-1)T_{\rm w}$ and $tT_{\rm w}$, where $t=1, \ldots, r$ is denoted $w_{ij}(t)$. The $N$ $\times$ $N$ adjacency matrix at time $t$ is given by $w(t)=(w_{ij}(t))$. We assume that links are undirected and thus the adjacency matrices are symmetric. However, the matrices may be weighted with link weights restricted to integers if multiple contacts are observed between two nodes during a single snapshot. The aggregate network (sometimes called the static network) is given by $\sum_{t=1}^r w(t)$.

\subsection{Transition probability}\label{sub:transition probability}

The definition of a transition matrix for temporal networks is nontrivial because it is necessary to determine the transition probability at isolated nodes. In general, some nodes may be isolated in a snapshot even if the aggregate network is connected. This is particularly the case when the time window for defining the snapshot, $T_{\rm w}$, is small. We thus assume that the random walker at an isolated node does not move in the corresponding snapshot. We also assume that the random walker does not move with some probability $q$ (i.e.\ the sojourn probability) if the node is not isolated. A similar idea of lazy random walks was introduced in Ref.~\cite{Ribeiro2013SciRep}, and the case $q=0$ was explored in Ref.~\cite{Starnini2012PhysRevE}. When $0<q<1$, the mixing property is guaranteed for general temporal networks such that an arbitrary initial distribution of random walkers converges to the unique stationary density (See more about mixing properties in Section~\ref{sub:mixing property}).

To define the transition probability, we start with the case in which nodes $i$ and $j$ are adjacent and they are not adjacent to any other node at time $t$. Then, we assume that in this snapshot a walker at $i$ moves to $j$ with probability $1-q$ and stays at $i$ with probability $q$. Similarly, a walker at $j$ moves to $i$ with probability $1-q$ and stays at $j$ with probability $q$. If other node pairs $i^{\prime}$ and $j^{\prime}$ are adjacent, and $i^{\prime}$ and $j^{\prime}$ are not adjacent to any other node at time $t$, the walkers transit between $i^{\prime}$ and $j^{\prime}$ with the same probabilities.

If $i$ is adjacent only to node $j_1$ at time $t$ and node $j_2$ at time $t+1$, the walker persists to node $i$ after two time steps with probability $q^2$. On the basis of this observation, we assume that the walker at $i$ does not move with probability $q^2$ in the snapshot in which $i$ is adjacent to $j_1$ and $j_2$. The walker moves to either $j_1$ or $j_2$ with probability $(1-q^2)/2$. It should be noted that the probability of the move to $j_1$ and $j_2$ is equal to $(1-q)$ and $q(1-q)$, respectively, when the two contacts ($i$, $j_1$) and ($i$, $j_2$) appear consecutively, not simultaneously.
In this case, the temporal order of the two contacts matters because $(1-q)>q(1-q)$. In contrast, the two probabilities are the same when nodes $j_1$ and $j_2$ are simultaneously adjacent to $i$.

In general, we define the transition probability from node $i$ to node $j$ at time $t$ as
\begin{equation}
B_{ij}(t)=\begin{cases}
\delta_{ij} &  (s_i(t)=0, 1\le j\le N),\\
q^{s_i(t)} & (s_i(t)\ge 1, i=j),\\
w_{ij}(t)(1-q^{s_i(t)})/s_i(t) & (s_i(t)\ge 1, i\neq j),
\end{cases}
\label{eq:def B}
\end{equation}
where $\delta_{ij}$ is the Kronecker delta and
\begin{equation}
s_i(t) \equiv \sum_{j=1}^N w_{ij}(t)
\end{equation}
is the node strength, i.e.\ the number of contacts that node $i$ has, at time $t$. Note that $\sum_{j=1}^N B_{ij}(t)=1$. The transition matrix at time $t$ is given by $B(t) = (B_{ij}(t))$.

We define an one-cycle transition matrix for the temporal (abbreviated as tp) network as
\begin{equation}
P^{\rm tp}\equiv\prod_{t=1}^r B(t).
\label{eq:P^{tp}}
\end{equation}
The stationary density of the random walk under the periodic boundary condition is given by the leading eigenvector (corresponding to the eigenvalue equal to unity) of $P^{\rm tp}$. The periodic boundary condition is given by the sequence $\ldots$, $w(1)$, $w(2)$, $\ldots$, $w(r)$, $w(1)$, $w(2)$, $\ldots$ and is necessary because of the finite observation time of an empirical temporal network~\cite{Starnini2012PhysRevE}.

When $q\approx 1$, Eq.~\eqref{eq:def B} is reduced to
\begin{equation}
B_{ij}(t)=\begin{cases}
\delta_{ij} & (s_i(t)=0, 1\le j\le N),\\
1- s_i(t)\epsilon & (s_i(t)\ge 1, i=j),\\
w_{ij}(t)\epsilon & (s_i(t)\ge 1, i\neq j),
\end{cases}
\label{eq:def B eps -> 0}
\end{equation}
up to the first order of $\epsilon\equiv 1-q\ll
1$. By combining Eqs.~\eqref{eq:P^{tp}} and
\eqref{eq:def B eps -> 0}, we obtain
\begin{equation}
P^{\rm tp}_{ij}=\begin{cases}
1- s_i^{\rm ag}\epsilon & (i=j),\\
\sum_{t=1}^r w_{ij}(t)\epsilon & (i\neq j),
\end{cases}
\label{eq:P^{tp} eps -> 0}
\end{equation}
where
\begin{equation}
s_i^{\rm ag}\equiv \sum_{t=1}^r s_i(t) = \sum_{t=1}^r \sum_{j=1}^N w_{ij}(t)
\label{eq:def s_i}
\end{equation}
is the node strength in the aggregate (abbreviated as ag in Eq.~\eqref{eq:def s_i}) network. Equation~\eqref{eq:P^{tp} eps -> 0} is the transition probability of the continuous-time random walk on the aggregate network for infinitesimally small time $\epsilon$.

\subsection{Mixing property}\label{sub:mixing property}

The mixing property, which guarantees the convergence to a unique stationary density starting from an arbitrary initial density, holds true for $0<q<1$, if and only if the aggregate network is connected. If the aggregate network is disconnected, trivially the random walk is not mixing. On the other hand, if the aggregate network is connected, there is a path of length $L_{ij}$ from any node $i$ to any node $j$ in the aggregate network. With a positive probability, a random walker located at $i$ travels on the first link of this path in a snapshot and does not move in all other snapshots in the first cycle of the application of $P^{\rm tp}$. Then, the random walker moves to the neighbor of $i$ on the mentioned path. Similarly, a random walker moves to a next node on the path in the second cycle with a positive probability, and so on. Therefore, the walker moves from $i$ to $j$ after $L_{ij}$ cycles with a positive probability. In addition, the walker moves from $i$ to $j$ after $L_{ij}+1$ cycles with a positive probability by never moving in one of the $L_{ij}+1$ cycles. Because $i$ and $j$ are arbitrary, $P^{\rm tp}$ is a positive matrix, i.e.\ any entry of $\left(P^{\rm tp}\right)^{\ell}$ for some integer $\ell (\ge 1)$ is positive, which renders the random walk mixing.

Nevertheless, if $q=0$, the mixing property is not necessarily satisfied even if the aggregate network is positive. In particular, if each node has at most one neighbor in each snapshot, the random walk is not mixing. For example, the adjacency matrix of the triangle is a positive matrix such that the random walk on the static triangle network is mixing. On the other hand, in the temporal network with $r=3$ in which each of the three snapshots contains just one contact, i.e.\ $w_{12}(1)=w_{21}(1)=w_{13}(2)=w_{31}(2)=w_{23}(3)=w_{32}(3)=1$, and all other $w_{ij}(t)=0$, the walker starting from node 1 comes back to node 1 with probability one at $t=3$. This means that the random walk on the temporal network is not mixing although the aggregate network of this temporal network is the triangle. In practice, this situation typically occurs when the temporal resolution of the data is high and $T_{\rm w}$ is small. If $q=0$, the density of the random walker depends on the initial density even in the limit $t\to\infty$. In particular, if the random walker starts from one node, the density is concentrated on a single node at any time because the random walker has only one option within each snapshot. Finally, if $q=1$, the walker never moves, and the random walk is not mixing

\subsection{Stationary density and the definition of the TempoRank}\label{sub:stationary density}

Assume that the random walk induced by $P^{\rm tp}$ is mixing.
We denote the unique stationary density of the random walk, i.e.\
the leading left eigenvector of $P^{\rm tp}$, by
\begin{equation}
\bm v(1) = \left(v_1(1) \; v_2(1) \; \cdots \; v_N(1)\right),
\end{equation}
where $v_i(1)$ ($1\le i\le N$) is the stationary density at node $i$.
In other words,
\begin{equation}
\bm v(1) = \bm v(1) P^{\rm tp}.
\label{eq:eigen equation}
\end{equation}
The normalization is given by $\sum_{i=1}^N v_i(1) = 1$.

In fact, $\bm v(1)$ is the stationary density when we observe the random
walk at $t=mr$, where $m$ is integer and tends to $\infty$. In general,
the density fluctuates even in the stationary state because we 
periodically apply different snapshots to move the walker.
For example, the stationary density
when we observe the random walk at $t=mr+1$, where $m\to\infty$, is
given by $\bm v(2) \equiv \bm v(1) B(1)$.
The long-term stationary density, i.e.\ that averaged within a cycle,
is given by
\begin{equation}
\overline{\bm v} \equiv \frac{1}{r} \sum_{t=1}^r \bm v(t),
\label{eq:def overline v}
\end{equation}
where
\begin{equation}
\bm v(t) = \bm v(1) \prod_{t^{\prime}=1}^{t-1}B(t^{\prime}).
\label{eq:def v(t)}
\end{equation}
We define $\overline{\bm v}=(\overline{v}_1\; \cdots\; \overline{v}_N)$ as the temporal random walk centrality, abbreviated as the TempoRank.

Similar to the case of the random walk on static networks, $\overline{v}_i$ is also interpreted as the total inflow to node $i$. In the stationary state, the inflow to node $i$ at time $t=mr+1$ is given by $(\bm v(1) B(1))_i = (\bm v(2))_i$ because $\bm v(1)$ is the stationary density at $t=mr$ and $B(1)$ is the transition matrix at $t=mr+1$. The inflow to node $i$ at $t=mr+2$ is given by $(\bm v(2) B(2)) = (\bm v(3))_i$ because $\bm v(2)$ is the stationary density at $t=mr+1$ and $B(2)$ is the transition matrix at $t=mr+2$. Same for $t=mr+3, \ldots, (m+1)r$. The total inflow of the probability to node $i$ in a cycle is given by $(\bm v(2))_i + (\bm v(3))_i + \cdots + (\bm v(r))_i + (\bm v(r)B(r))_i = (\bm v(2))_i + (\bm v(3))_i + \cdots + (\bm v(r))_i + (\bm v(1))_i = r\overline{v}_i$. Therefore, $\overline{v}_i$ is equal to the average inflow to node $i$ per time step.

The stationary densities $\bm v(1)$ and $\overline{\bm v}$ depend on the value of $q$, which contrasts to the results obtained from a different model \cite{Ribeiro2013SciRep}.
The temporal random walk induced by $P^{\rm tp}$ coincides with the continuous-time random walk in the aggregate network in the limit $q\to 1$ (Eq.~\eqref{eq:P^{tp} eps -> 0}). In general, the stationary density of the continuous-time random walk in a connected network is given by $1/N$ at each node \cite{Lambiotte2008arxiv}.  Therefore, we obtain $\bm v(t)$ $(1\le t\le r)$, $\overline{\bm v}\to (1\; \cdots\; 1)/N$ in the limit $q\to 1$.

\subsection{Random walk on the aggregate network}

The transition matrix of the discrete-time random walk on the aggregate network is given by
\begin{equation}
P^{\rm ag} \equiv \left[ \sum_{t=1}^r D(t)\right]^{-1}
\left[\sum_{t=1}^r w(t)\right],
\label{eq:P^{ag}}
\end{equation}
where the $N$ $\times$ $N$ diagonal matrix $D(t)$ is defined by $D_{ij}(t)=\delta_{ij}\sum_{j=1}^N w_{ij}(t)$ $(=\delta_{ij}\sum_{j=1}^N w_{ji}(t))$. The diagonal elements of $\sum_{t=1}^r D(t)$ are equal to the node strength of the aggregate network given by Eq.~\eqref{eq:def s_i}.

$P^{\rm ag}$ is distinct from $P^{\rm tp}$ or its weighted versions.
For example, $P^{\rm tp}_{ij}>0$ if there is a temporal path from $i$ to $j$ whose length is at most $r$, whereas $P^{\rm ag}_{ij}>0$ ($i\neq j$) if and only if $i$ and $j$ are adjacent.

\section{The effective network and the in-strength approximation}

In this section, we show that the TempoRank is equal to the stationary density of the discrete-time random walk on a static weighted and directed network, which we call the effective network. In other words, we map the random walk on a temporal network into a directed weighted static network. This relationship allows us to give a new interpretation to the TempoRank and to develop a local approximator.

Under $0\le q<1$, Eq.~\eqref{eq:def B} is equivalent to
\begin{equation}
B_{ij}(t) = \frac{w^{\prime}_{ij}(t)}{\sum_{\ell=1}^N w^{\prime}_{i\ell}(t)},
\end{equation}
where
\begin{equation}
w^{\prime}_{ij}(t)=
\begin{cases}
\delta_{ij} & (s_i(t)=0,  1\le j\le N),\\
s_i(t)q^{s_i(t)}/(1-q^{s_i(t)}) & (s_i(t)\ge 1, i=j),\\
w_{ij}(t) & (s_i(t)\ge 1, i\neq j).
\end{cases}
\end{equation}
In terms of the undirected weighted matrix
$w^{\prime}(t)=(w^{\prime}_{ij}(t))$, we obtain

\begin{align}
P^{\rm tp}_{ij} =& \sum_{k_1,\ldots,k_{r-1}} \frac{w^{\prime}_{ik_1}(1)}{\sum_{\ell_1}w^{\prime}_{i\ell_1}(1)} \frac{w^{\prime}_{k_1 k_2}(2)}{\sum_{\ell_2} w^{\prime}_{k_1 \ell_2}(2)}
\cdots \frac{w^{\prime}_{k_{r-1}j}(r)}{\sum_{\ell_r} w^{\prime}_{k_{r-1}\ell_r}(r)}\notag\\
=& \frac{w^{\rm tp}_{ij}(1)}{\sum_{\ell}w^{\rm tp}_{i\ell}(1)},
\label{eq:P^{tp}_{ij}}\\
P^{\rm tp}_{ji} =& \sum_{k_1,\ldots,k_{r-1}} \frac{w^{\prime}_{jk_1}(1)}{\sum_{\ell_1}w^{\prime}_{j\ell_1}(1)} \frac{w^{\prime}_{k_1 k_2}(2)}{\sum_{\ell_2} w^{\prime}_{k_1\ell_2}(2)}\cdots
\frac{w^{\prime}_{k_{r-1}i}(r)}{\sum_{\ell_r} w^{\prime}_{k_{r-1}\ell_r}(r)}\notag\\
=& \frac{w^{\rm tp}_{ji}(1)}{\sum_{\ell}w^{\rm tp}_{j\ell}(1)},
\label{eq:P^{tp}_{ji}}
\end{align}
where
\begin{align}
w^{\rm tp}_{ij}(1) \equiv& \sum_{k_1,\ldots,k_{r-1}}
\frac{w^{\prime}_{ik_1}(1) w^{\prime}_{k_1k_2}(2)\cdots w^{\prime}_{k_{r-1}j}(r)}
{\sum_{\ell_1} w^{\prime}_{i\ell_1}(1) \sum_{\ell_2} w^{\prime}_{k_1\ell_2}(2)
\cdots \sum_{\ell_r}w^{\prime}_{k_{r-1}\ell_r}(r)},\\
w^{\rm tp}_{ji}(1) \equiv& \sum_{k_1,\ldots,k_{r-1}}
\frac{w^{\prime}_{jk_1}(1) w^{\prime}_{k_1k_2}(2)\cdots w^{\prime}_{k_{r-1}i}(r)}
{\sum_{\ell_1} w^{\prime}_{j\ell_1}(1) \sum_{\ell_2} w^{\prime}_{k_1\ell_2}(2)
\cdots \sum_{\ell_r}w^{\prime}_{k_{r-1}\ell_r}(r)}.
\end{align}
In fact,
\begin{equation}
\sum_{j=1}^N w^{\rm tp}_{ij}(1) =1
\label{eq:w^{tp} normalization}
\end{equation}
holds true for each $i$ such that
the denominators of the right-hand sides of Eqs.~\eqref{eq:P^{tp}_{ij}}
and \eqref{eq:P^{tp}_{ji}} are equal to unity.

Equations~\eqref{eq:P^{tp}_{ij}}
and \eqref{eq:P^{tp}_{ji}} indicate that
$P^{\rm tp}$ is the transition matrix of the discrete-time random walk on 
the static weighted network defined by $w^{\rm tp}(1)= (w^{\rm tp}_{ij}(1))$. We call this network the effective network.
In general,
\begin{equation}
w^{\rm tp}_{ij}(1)\neq w^{\rm tp}_{ji}(1).
\label{eq:w^{tp} asymmetric}
\end{equation}
Each snapshot $w^{\prime}(t)$ and the aggregate network are undirected networks. However, the concatenation of the different snapshots makes the effective network directed due to the arrow of time, which creates asymmetry in the sequence of link activation.

For static directed networks, the in-degree is often accurate in
approximating the stationary density of the random walk
\cite{Amento2000ACM_SIGIR,Nakamura2003PhysRevE,Fortunato2008LNCS,MasudaOhtsuki2009NewJPhys,Ghoshal2011NatComm}.
%
%
Here we develop the same type of local approximation of the TempoRank
by considering the in-strength of nodes in the effective network.
The in-degree of node $i$ does not depend on the out-degree of the upstream neighbors of $i$. In contrast, the in-strength
is large (small) when the out-degree of an upstream neighbor is
small (large). The in-strength
of node $i$ for the effective network is given by
\begin{equation}
s_i^{\rm tp}(1) \equiv \sum_{j=1}^N w^{\rm tp}_{ji}(1).
\label{eq:s_i^{tp}(1)}
\end{equation}
The in-strength is considered to be an appropriate approximator to the stationary density because Eqs.~\eqref{eq:eigen equation}, \eqref{eq:P^{tp}_{ji}},
and \eqref{eq:w^{tp} normalization} imply
\begin{equation}
v_i(1) = \sum_{j=1}^N v_j(1) w^{\rm tp}_{ji}(1) \approx \left<v(1)\right>
\sum_{j=1}^N w^{\rm tp}_{ji}(1) = \frac{s_i^{\rm tp}(1)}{N},
\label{eq:reason for in-strength approximator}
\end{equation}
if all $v_j(1)$'s are approximated by an ensemble average given by $\left<v(1)\right>=1/N$. Equations~\eqref{eq:w^{tp} asymmetric} and \eqref{eq:s_i^{tp}(1)} imply $\sum_{i=1}^N s_i^{\rm tp}(1) = N$. Therefore, Equation~\eqref{eq:reason for in-strength approximator} provides a normalized in-strength approximator to $\bm v(1)$.

We calculate the in-strength approximator to $\overline{\bm v}$ as follows. Because $\bm v(1) P^{\rm tp} = \bm v(1) B(1)B(2)\cdots B(r) = \bm v(1)$ implies that
$\bm v(t) B(t) B(t+1) \cdots B(r)B(1)\cdots $ $B(t-1)
= \bm v(t)$ ($2\le t\le r$),
$\bm v(t)$ is the stationary density of the random walk in 
the static network $w^{\rm tp}(t) = (w_{ij}^{\rm tp}(t))$ defined by
\begin{align}
w^{\rm tp}_{ij}(t) \equiv& \sum_{k_1,\ldots,k_{r-1}}
\frac{w^{\prime}_{ik_1}(t) w^{\prime}_{k_1k_2}(t+1)\cdots w^{\prime}_{k_{r-1}j}(t-1)}
{\sum_{\ell_1} w^{\prime}_{i\ell_1}(t) \sum_{\ell_2} w^{\prime}_{k_1\ell_2}(t+1)
\cdots \sum_{\ell_r}w^{\prime}_{k_{r-1}\ell_r}(t-1)},\\
w^{\rm tp}_{ji}(t) \equiv& \sum_{k_1,\ldots,k_{r-1}}
\frac{w^{\prime}_{jk_1}(t) w^{\prime}_{k_1k_2}(t+1)\cdots w^{\prime}_{k_{r-1}i}(t-1)}
{\sum_{\ell_1} w^{\prime}_{j\ell_1}(t) \sum_{\ell_2} w^{\prime}_{k_1\ell_2}(t+1)
\cdots \sum_{\ell_r}w^{\prime}_{k_{r-1}\ell_r}(t-1)}.
\end{align}
Therefore, the in-strength approximation is given by
\begin{equation}
\overline{v}_i \approx \frac{\sum_{t=1}^r s_i^{\rm tp}(t)}{Nr},
\label{eq:normalized approximator overline v}
\end{equation}
where
\begin{equation}
s_i^{\rm tp}(t) \equiv \sum_{j=1}^N w^{\rm tp}_{ji}(t).
\label{eq:s_i^{tp}(t)}
\end{equation}

\section{Numerical analysis}\label{sec:numerical results}

In this section, we numerically examine the TempoRank. We assess the performance of the in-strength approximation on empirical temporal networks and discuss the right moment hypothesis, i.e.\ the contention that nodes have to contact other nodes at the right moment.

\subsection{Data sets}

We performed the numerical analysis using the following empirical networks. One network represents face-to-face interactions between conference attendees (SPC)~\cite{Isella11}, another corresponds to the same type of interactions between visitors of a museum (SPM)~\cite{Isella11}, and the third to proximity between staff and patients in a hospital (SPH)~\cite{Vanhems13}. The fourth data set corresponds to sexual contacts between sex-sellers and -buyers extracted from a webforum (SEX)~\cite{Rocha2010PNAS}. The last data set is a sample of email communication between students and staff within a university (EMA)~\cite{Eckmann2004PNAS}. These networks represent human interactions in diverse social contexts and have different topological and temporal characteristics (Table~\ref{tab_1}).

\begin{table}[htb]
\begin{center}
\caption{Summary information about the empirical networks. Number of nodes ($N$), number of links ($|E|=\sum_i s_i^{\rm ag}/2$), recording time ($T$), and maximum temporal resolution ($\delta$).}
\begin{tabular}{cccccc}
\hline
       & $N$ & $|E|$ & $T$ (days) & $\delta$ \\
\hline
SPC & 113 & 20,818 & $\sim 2.5$ & 20 sec \\
SPM & 72 & 6,980 & $\sim 1$ & 20 sec \\
SPH & 75 & 32,424 & $\sim 4$ & 20 sec \\
SEX & 1,302 & 1,814 & 50 & 1 day \\
EMA & 1,564 & 4,461 & 1 & 1 sec \\
\hline
\end{tabular}
\label{tab_1}
\end{center}
\end{table}

\subsection{Numerical procedures}

Consider a given sequence of snapshots $w(1)$, $\ldots$, $w(r)$. We apply the power method on $P^{\rm tp}$ (Eq.~\eqref{eq:P^{tp}}) to obtain the stationary density $\bm v(1)$ and use Eqs.~\eqref{eq:def overline v} and \eqref{eq:def v(t)} to calculate the TempoRank $\overline{\bm v}$. The initial condition is the uniform density $\bm{v}_{\rm init}(1) = (1\; \cdots \; 1)/N$, and iteration stops when $\sqrt{(\bm{v}_{\rm post}(1) -\bm{v}_{\rm pre}(1))^2/N} < 10^{-6}$ for the first time, where $\bm{v}_{\rm pre}(1)$ and $\bm{v}_{\rm post}(1)$ are the estimation of $\bm v(1)$ before and after multiplying $P^{\rm tp}$, respectively, during the power iteration. The stationary density for the aggregate network, i.e.\ the normalized leading eigenvector of the transition matrix $P^{\rm ag}$ (Eq.~\eqref{eq:P^{ag}}), is given by the (normalized) strength of the node in the aggregate network (Eq.~\eqref{eq:def s_i}). This is because the network is undirected~\cite{Doyle1984book,Lovasz1993Boyal}.

\subsection{In-strength approximation}\label{sub:in-strength approximation}

\begin{figure*}[htb]
\centering
\includegraphics[scale=0.8]{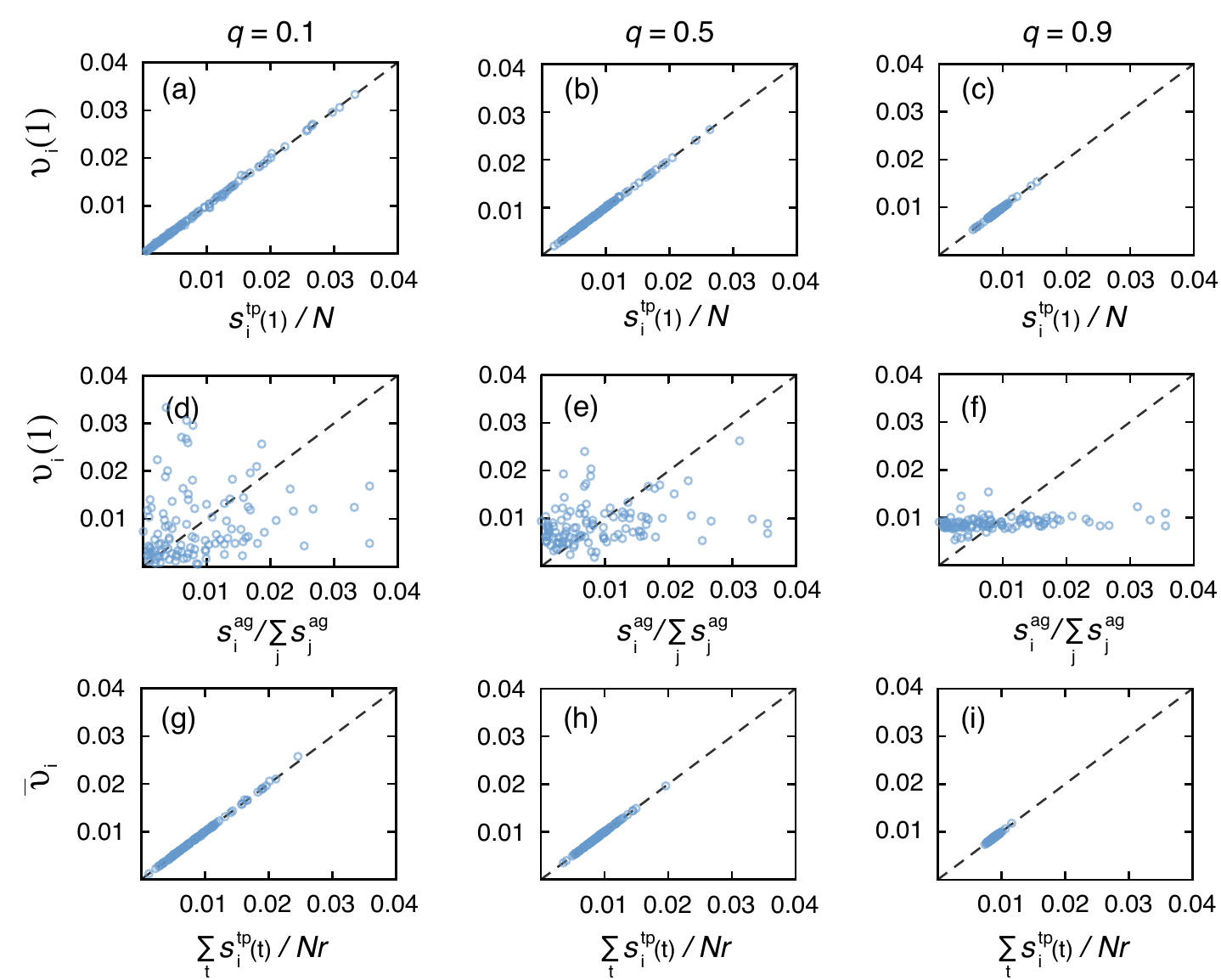}
\caption{\textbf{TempoRank for SPC network.} (a)--(c) Relationship between the stationary density of the temporal transition matrix and the in-strength approximation. Each circle represents a node, and the dashed lines represent the diagonal. (d)--(f) Relationship between the stationary density of the temporal transition matrix and the in-strength of the aggregate network. (g)--(i) Relationship between the TempoRank and the time-averaged in-strength of the effective network. We set $q=0.1$ in (a), (d), (g), $q=0.5$ in (b), (e), (h), and $q=0.9$ in (c), (f), (i). The resolution is $T_{\rm w}= 5$ min.}
\label{fig1}
\end{figure*}

Figure~\ref{fig1}(a)--(c) shows the performance of the in-strength approximator with three values of $q$ for the SPC data set. The in-strength approximation is accurate for a wide range of values of $q$ (i.e.\ from 0.1 to 0.9) for most nodes. In contrast, the in-strength of the aggregate network, i.e.\ $s_i^{\rm ag}$, which gives the exact stationary density of the random walk on the aggregate network, is little correlated with $v_i(1)$ for the same three values of $q$ (Fig.~\ref{fig1}(d)--(f)). The in-strength approximator for the TempoRank (Eq.~\eqref{eq:normalized approximator overline v}) is also strongly correlated with $\overline{v}_i$, as shown in Fig.~\ref{fig1}(g)--(i). The results are qualitatively the same for the other empirical networks, as shown in Fig.~\ref{fig2} (for SPM data set) and in Supplementary Material (for the other data sets).

The values of $v_i(1)$ and $\overline{v}_i$ are similar for all nodes when $q=0.9$ (Figs.~\ref{fig1}(c),~\ref{fig1}(f), and~\ref{fig1}(i)). This result is consistent with the theoretical prediction made in section~\ref{sub:stationary density}, i.e.\ $\bm v(1), \overline{\bm v}\to (1\; \cdots\; 1)/N$ as $q\to 1$.

\begin{figure*}[htb]
\centering
\includegraphics[scale=0.8]{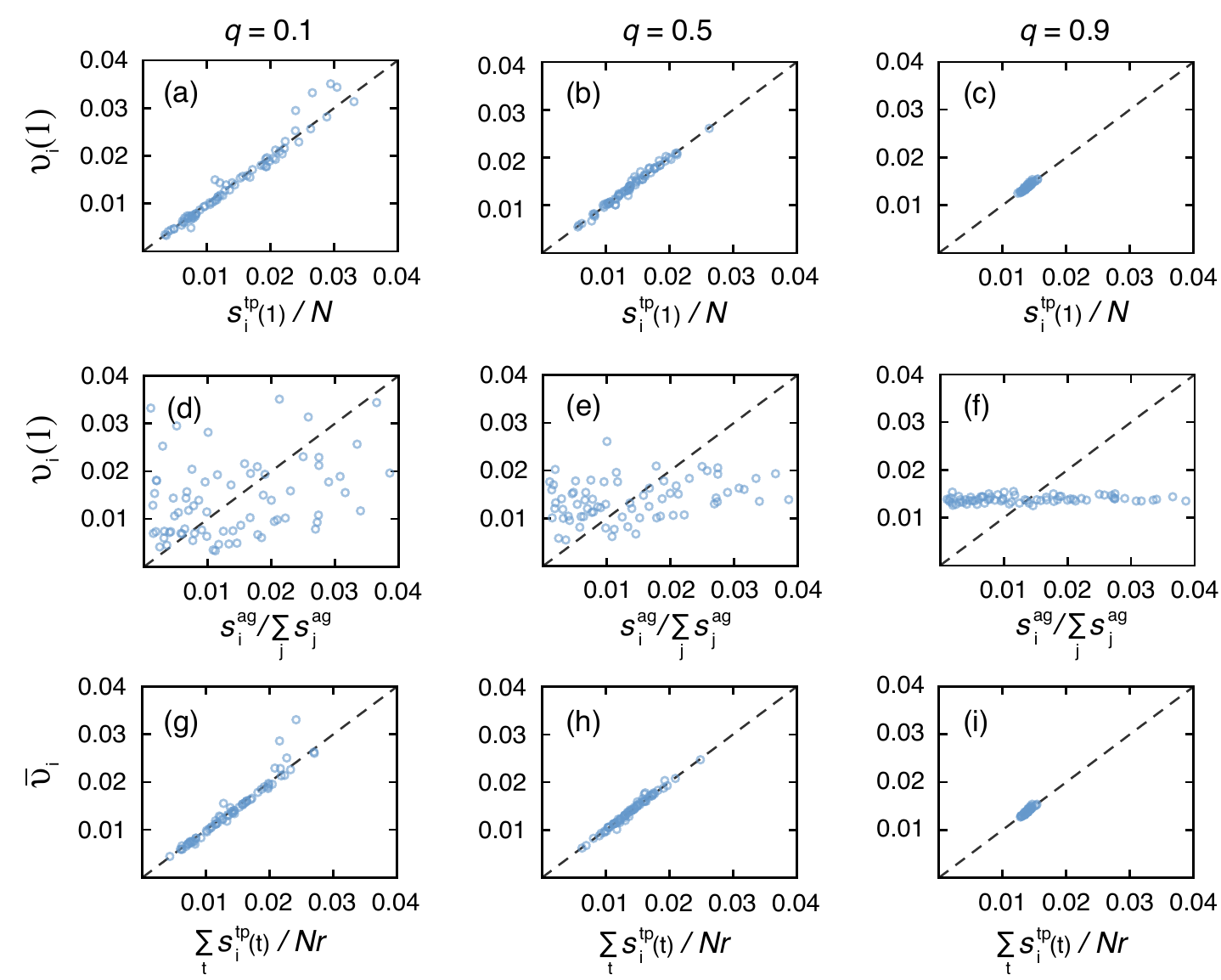}
\caption{\textbf{TempoRank for SPM network.} (a)--(c) Relationship between the stationary density of the temporal transition matrix and the in-strength approximation. (d)--(f) Relationship between the stationary density of the temporal transition matrix and the in-strength of the aggregate network. (g)--(i) Relationship between the TempoRank and the time-averaged in-strength of the effective network. We set $q=0.1$ in (a), (d), (g), $q=0.5$ in (b), (e), (h), and $q=0.9$ in (c), (f), (i). The resolution is $T_{\rm w}= 1$ min.}
\label{fig2}
\end{figure*}

\subsection{The right-moment hypothesis}

In principle, the stationary density of the random walk at a node is large if the node receives links from nodes with high stationary densities. This principle underlies the design of the PageRank~\cite{Brin1998conf,Langville2006book}. More generally, the principle that being adjacent to a central node is important, for the node itself to be important, guides the definition of the Katz centrality, eigenvector centrality, and their variants~\cite{Newman2010book}. In the case of the TempoRank, however, we show in section~\ref{sub:in-strength approximation} that the in-strength approximation for the effective network, which ignores the ``next-to-celebrity'' principle, is pretty accurate.

The ``next-to-celebrity'' principle accommodated to the TempoRank dictates that a node $i$ being connected to another node with a large density of walkers at the right moment gains a large inflow at that time, leading to a large $\overline{v}_i$ value. Some centrality measures for temporal networks on the basis of the right-moment principle have been proposed in different forms~\cite{Motegi2012SciRep,Grindrod2013SiamRev}. Our results in section~\ref{sub:in-strength approximation} implies that the right-moment principle is practically irrelevant in the TempoRank.

\begin{figure*}[htb]
\centering
\includegraphics[scale=0.8]{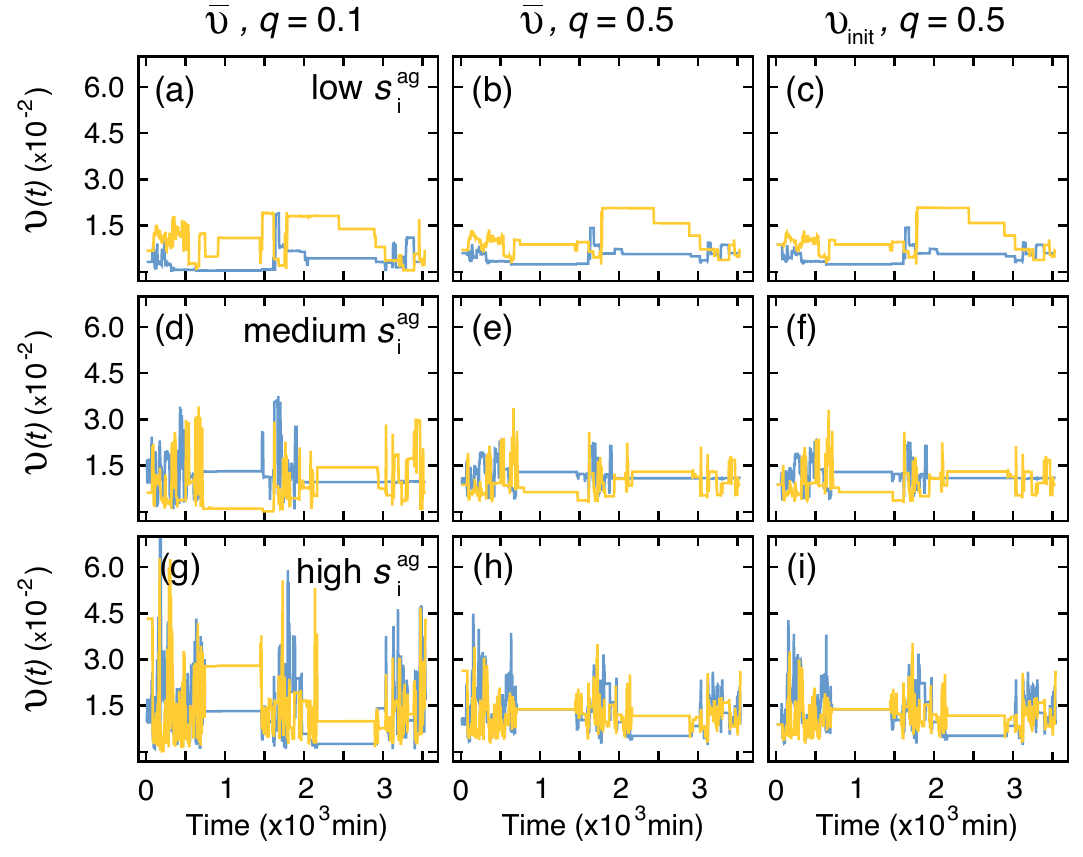}
\caption{\textbf{Time dependence of the stationary density of the random walk for the SPC data set.} In (a), (b), (d), (e), (g), and (h), the density of walkers $\bm v(t)=\overline{\bm v} B(1)\cdots B(t-1)$ is shown. In (c), (f), and (i), the density of walkers in a snapshot $t$ calculated by $\bm v(t) = \bm{v}_{\rm init}(1) B(1)\cdots B(t-1)$, where $\bm{v}_{\rm init}(1) = (1\; \cdots\; 1)/N$, is shown. Each curve corresponds to a node with different $s^{\rm ag}_i$; the two curves in each panel represents two representative nodes in the corresponding node-strength category. We set $q=0.1$ in (a), (d), (g), and $q=0.5$ in (b), (c), (e), (f), (h), (i). The resolution is $T_{\rm w}= 5$ min.}
\label{fig:SPC T_w=5}
\end{figure*}

To examine this point, we measure the temporal fluctuation of the density of random walkers within a cycle. For the SPC data set, the stationary density at the $t$-th snapshot, i.e.\ $v_i(t)$, for six nodes is shown as a function of $t$ in Figs.~\ref{fig:SPC T_w=5}(a), (d), (g) with $q=0.1$ and Figs.~\ref{fig:SPC T_w=5}(b), (e), (h) with $q=0.5$. Each panel represents the time course of the stationary density for two representative nodes $i$ with low, intermediate, and high strengths in the aggregate network, i.e.\ the total number of contacts, $s_i^{\rm ag}$. The corresponding results for the SPM data set are shown in Fig.~\ref{fig:SPM T_w=1}. The results for the other data sets are shown in Supplementary Material. Figures~\ref{fig:SPC T_w=5} and \ref{fig:SPM T_w=1} indicate that the fluctuation of $v_i(t)$ is large irrespective of the strength of the node, although it is larger for nodes with larger strength values. The density of walkers remains constant at times when nodes are not making contacts. In particular, we identify plateaus for some ranges of $t$, which correspond to night periods in case of the SPC network (Fig.~\ref{fig:SPC T_w=5}) and SPH network (see Supplementary Material). In the SPM network (Fig.~\ref{fig:SPM T_w=1}), most plateaus correspond to earlier or later times because visits are organized in groups in this museum, allowing interactions only within limited time windows. The fluctuations decrease as $q$ increases. This behavior is expected because the stationary density approaches the uniform density as $q$ increases.

\begin{figure*}[htb]
\centering
\includegraphics[scale=0.8]{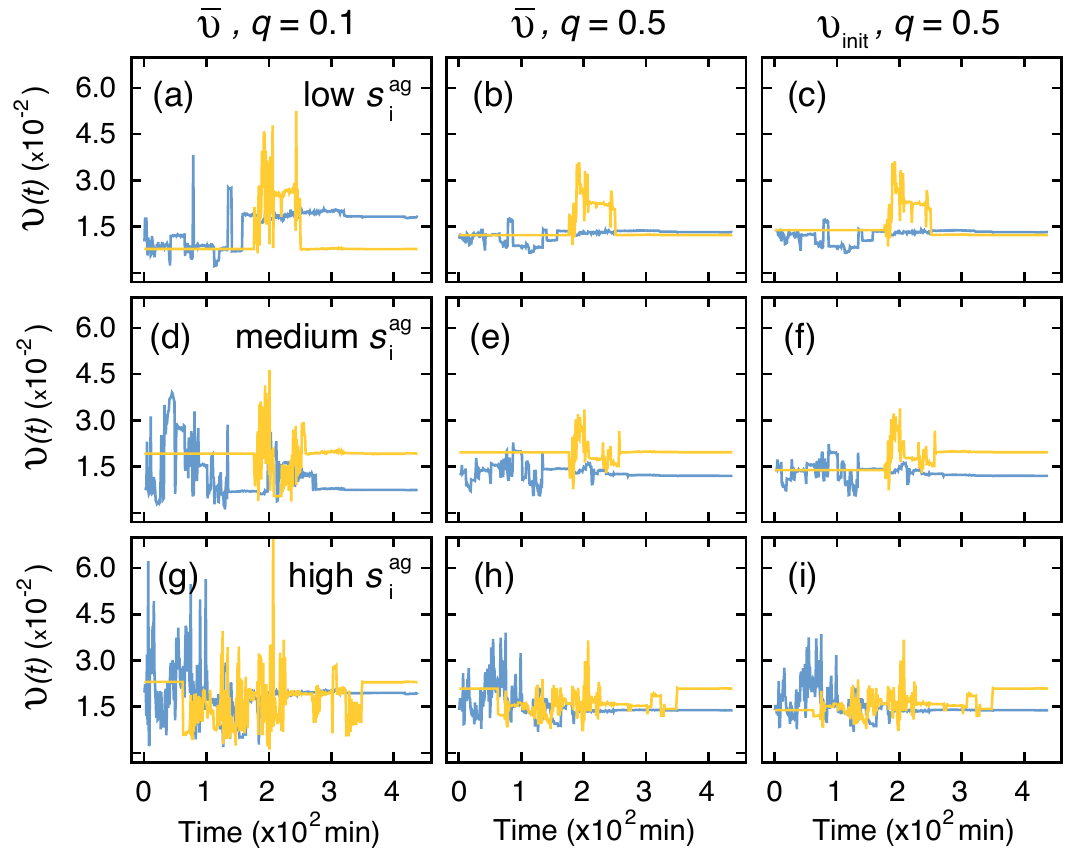}
\caption{\textbf{Time dependence of the stationary density of the random walk
for the SPM data set.} 
The resolution is $T_{\rm w}= 1$ min.
See the legends of Fig.~\ref{fig:SPC T_w=5} for other details.}
\label{fig:SPM T_w=1}
\end{figure*}

\begin{figure*}[htb]
\centering
\includegraphics[scale=0.8]{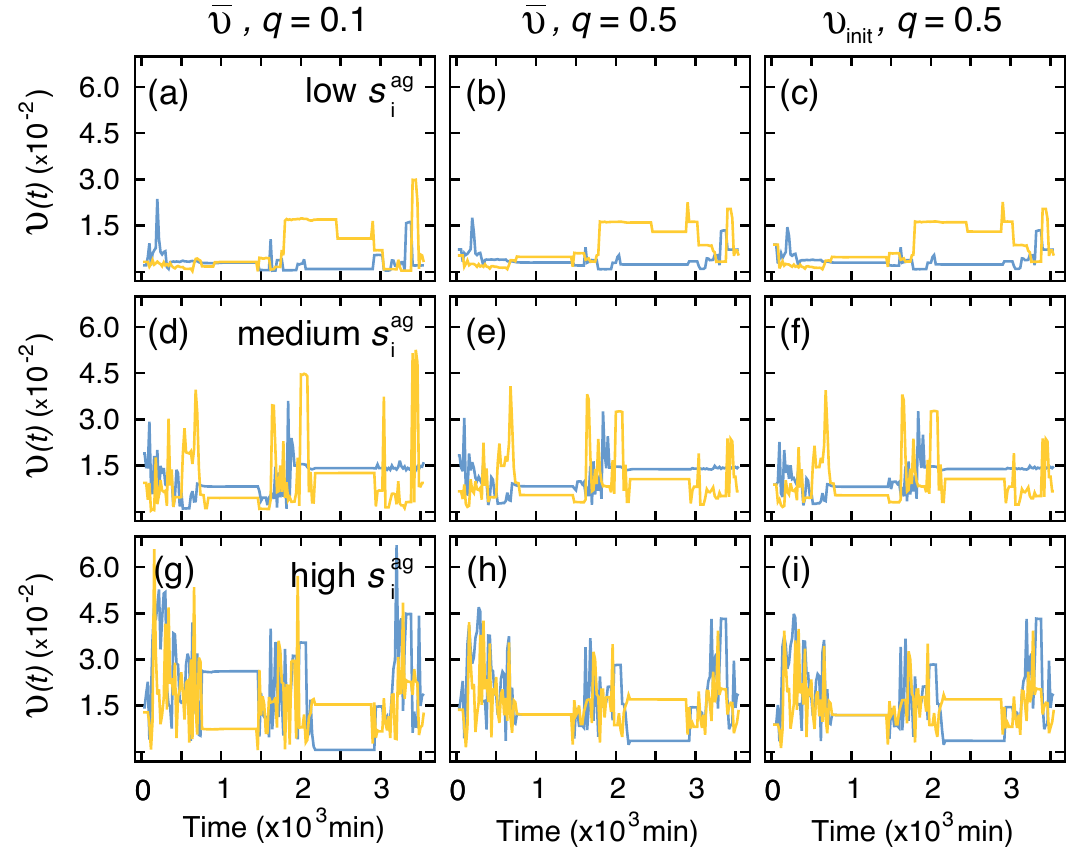}
\caption{\textbf{Time dependence of the stationary density for the SPC data set with a lower resolution.}  
The resolution is $T_{\rm w}= 20$ min.
See the legends of Fig.~\ref{fig:SPC T_w=5} for other details.}
\label{fig:SPC T_w=20}
\end{figure*}

\begin{figure*}[htb]
\centering
\includegraphics[scale=0.8]{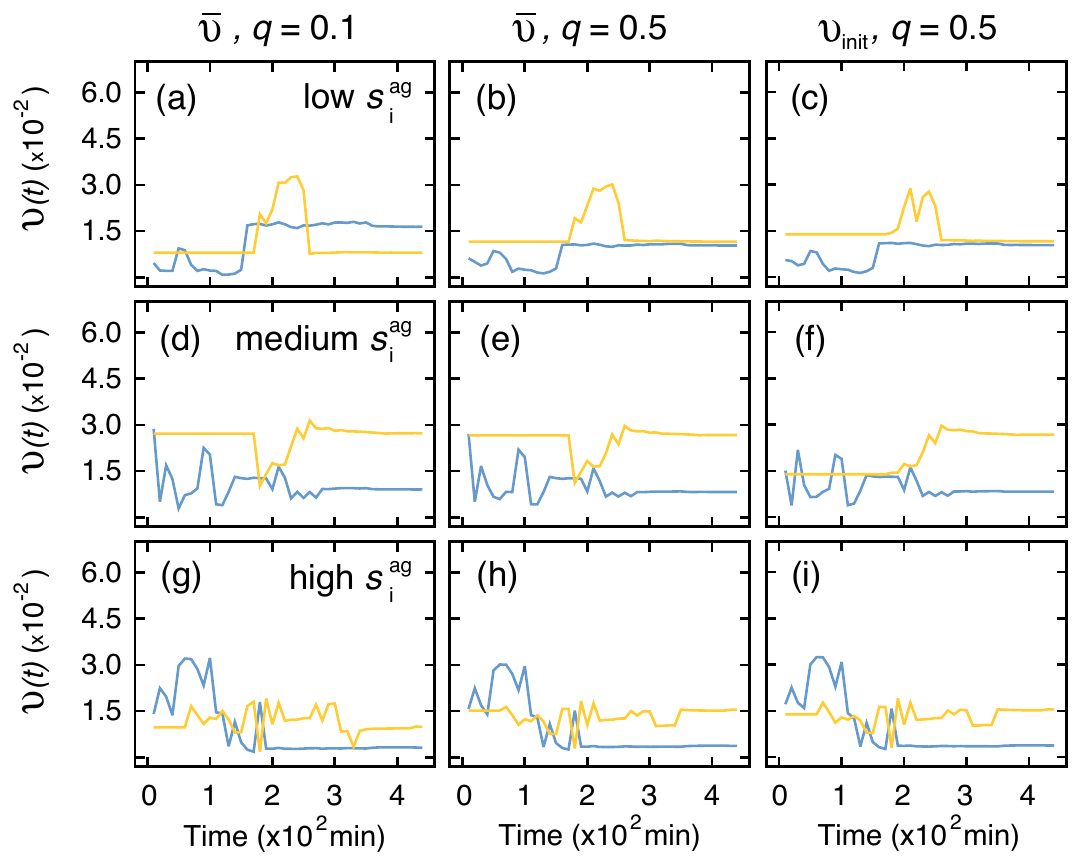}
\caption{\textbf{Time dependence of the stationary density for the SPM
data set with a lower resolution.} The resolution is $T_{\rm w}= 10$ min.
See the legends of Fig.~\ref{fig:SPC T_w=5} for other details.}
\label{fig:SPM T_w=10}
\end{figure*}

Similar temporal fluctuations are observed when the temporal networks are coarse-grained (i.e.\ with a low resolution), as shown in Figs.~\ref{fig:SPC T_w=20} and~\ref{fig:SPM T_w=10} for the SPC and SPM data sets, respectively. The same is valid for the other data sets (see Supplementary Material). Therefore, large fluctuations are a general phenomenon irrespective of the temporal resolution.

The large fluctuations revealed in Figs.~\ref{fig:SPC T_w=5}--\ref{fig:SPM T_w=10} suggest that a node should be adjacent to nodes with high density of walkers at the right moment to secure a large $\overline{v}_i$ value, in favor of the right-moment principle. Nevertheless, the high accuracy of the in-strength approximation does not support the relevance of the right-moment principle.

We can resolve this apparent paradox as follows. The in-strength in the effective network, $s_i^{\rm tp}(1)$, consists of the contributions from different neighbors (i.e.\ $j$'s in Eq.~\eqref{eq:s_i^{tp}(1)}). Each $w^{\rm tp}_{ji}(1)$ is equal to the number of temporal paths from $j$ to $i$ in the one cycle starting and ending at $t=1$ and $t=r$, respectively. Each path is weighted by the out-degree of the source nodes on the path. If the out-degree of $j$ is large at $t=1$, the flow of the random walk is equally divided by the downstream neighbors such that a downstream neighbor of $j$, denoted by $k_1$, receives a relatively small inflow of the random walk. Then, at $t=2$, node $k_1$, which has received the inflow of probability from its upstream neighbors (including $j$) at $t=1$, sends the flow to $k_1$'s downstream neighbors at $t=2$. If the number of neighbors is large, then each downstream neighbor of $k_1$ at $t=2$ receives a small inflow. Finally, $s_i^{\rm tp}(1)$ is the total inflow, or weighted path count summed over all the starting nodes $j$ at $t=1$. The crucial observation here is that in the in-strength definition, the starting node is not weighted. In contrast, the exact calculation of $v_i(1)$ assumes that the starting node is weighted according to the stationary density $v_j(1)$, as indicated in the first equality in Eq.~\eqref{eq:reason for in-strength approximator}.

Therefore, the fact that the in-strength approximation works well implies that the fluctuation of the density of walkers within a cycle starting from the uniform density or that starting from the stationary density do not significantly differ. This is in fact observed. In Figs.~\ref{fig:SPC T_w=5}(c), \ref{fig:SPC T_w=5}(f), \ref{fig:SPC T_w=5}(i) \ref{fig:SPM T_w=1}(c), \ref{fig:SPM T_w=1}(f), and \ref{fig:SPM T_w=1}(i), we show the fluctuation of the density of walkers starting from the uniform density, i.e.\ $(1\; \cdots\; 1)/N$ for the same selected nodes as those in Figs.~\ref{fig:SPC T_w=5}(b), \ref{fig:SPC T_w=5}(e), \ref{fig:SPC T_w=5}(h), \ref{fig:SPM T_w=1}(b), \ref{fig:SPM T_w=1}(e), and \ref{fig:SPM T_w=1}(h) (which correspond to the initial condition $\bm v(1)$). The fluctuation is similar between the two initial conditions except in early snapshots. Therefore, we conclude that the right-moment principle is logically present but practically unimportant.

\section{Discussion}

We proposed the TempoRank, a node centrality measure for temporal networks. In addition to the exact computation, we showed that the TempoRank is accurately approximated by the in-strength of the node in the effective network. The effective network is a directed network induced by the undirected temporal network. The concept of the effective network may be useful for other purposes, such as path counting of temporal networks and revealing information or viral flow along the arrow of time.

In static directed networks, the stationary density of the random walk often deviate substantially from the in-degree~\cite{Donato2004EPJB,MasudaOhtsuki2009NewJPhys, Volkovich2009LNICST}, whereas it is accurate in other cases~\cite{Amento2000ACM_SIGIR,Nakamura2003PhysRevE,Fortunato2008LNCS,MasudaOhtsuki2009NewJPhys, Ghoshal2011NatComm,Volkovich2009LNICST}. We found that the in-strength of the effective network approximates the TempoRank, i.e.\ the stationary density in the effective (directed) network, with a high accuracy. There are at least two possible reasons underlying the high accuracy of the in-strength approximation. 

First, the effective network is usually dense. In general, if there is a directed temporal path from node $i$ to node $j$ in the given temporal network, $w^{\rm tp}_{ij}\ge 1$. Therefore, the link density in the effective network is equal to the so-called reachability measure \cite{Holme2005PhysRevE}, except for the difference in the treatment of the diagonal elements $w^{\rm tp}_{ii}$. In many temporal network data sets, the reachability is moderately or very large even if each snapshot in the temporal network is sparse~\cite{Holme2005PhysRevE, Pan2011PhysRevE,Takaguchi2013PlosOne, Lentz2013PhysRevLett} unless the number of snapshots (i.e.\ $r$) is too small. Then, the effective networks are dense. In this situation, the summation is taken over many upstream neighbors of node $i$ for calculating $v_i(1)$ (first equality in Eq.~\eqref{eq:reason for in-strength approximator}). Then, the heterogeneity in the TempoRank among the upstream neighbors of $i$, because of which the in-strength may deviate from $v_i(1)$ (Eq.~\eqref{eq:reason for in-strength approximator}), may efficiently cancel out to yield similarity between the in-strength and $v_i(1)$. 

Second, the in-strength may be a significantly better approximator than the in-degree in general static and temporal networks. In the random walk on model temporal networks, the stationary density is only weakly correlated with the degree of the aggregate network~\cite{Perra2012PhysRevLett}. Investigating the performance of the in-strength approximator in this situation and also on static networks may be an interesting research question.

We assumed the periodic boundary condition in time to define the stationary density of the random walk. In fact, a real temporal network data set does not repeat itself; the first snapshot does not follow the last snapshot. In addition, temporal network data are often non-stationary, swamped by frequent overturns of nodes and links even within a recording period~\cite{Rocha2013PlosComputBiol,Holme2013PlosComputBiol,Rocha2013PhysRevE}. A justification of the use of the periodic boundary condition is that the convergence of the power iteration seems to be very fast unless the number of snapshots is small. This was observed when we started from different initial conditions to have almost the same density of walkers at various nodes after a short transient within a single cycle (comparison between panels (b), (e), (h) and panels (c), (f), (i) in Figs.~\ref{fig:SPC T_w=5}--\ref{fig:SPM T_w=10}). Therefore, the TempoRank represents the probability flow as we sequentially apply the snapshots in a single cycle. Investigating the generalizability of this result warrants future work. 

The diffusive dynamics in the continuous time is described by Laplacian dynamics. The Laplacian dynamics driven the unnormalized Laplacian matrix has the uniform stationary density both for the temporal network represented by succession of snapshots and aggregate network \cite{Masuda2013PhysRevLett}. In contrast, we showed that the stationary density differed between the temporal and aggregate networks when the diffusive dynamics was considered in discrete time. A lesson drawn from this consideration is that we should be careful in discrete versus continuous time when considering diffusive processes on temporal networks. Analyzing a continuous-time counterpart of the TempoRank or the stationary density, as touched upon in Ref.~\cite{Hoffmann2013chapter}, requests further studies.

To assure the mixing property in arbitrary connected temporal networks, we assumed that the walker resided in the current node with probability $q$. The original PageRank employs the so-called teleportation probability to make the random walk mixing for arbitrary static networks~\cite{Brin1998conf,Langville2006book}. The sojourn probability $q$, however, is unrelated to the teleportation probability. The latter dictates that a walker jumps to an arbitrary node with a given equal probability irrespective of the current position, while $q$ specifies the laziness of the random walk to move, as assumed in~\cite{Ribeiro2013SciRep}. In our model, a random global jump probability is unnecessary because the initial network is assumed to be undirected and periodic boundary conditions are adopted, which removes the possibility that walkers are trapped on certain nodes. Furthermore, the teleportation adds another hyperparameter (i.e.\ the teleportation probability) and blurs the effect of the original network because it is a network-independent random jump.

\section*{Acknowledgements}

We acknowledge Jean-Pierre Eckmann for kindly providing to us the data used in \cite{Eckmann2004PNAS}. We thank Ryosuke Nishi and Taro Takaguchi for careful reading of the manuscript. LECR is a Charg\'e de recherches of the Fonds de la Recherche Scientifique - FNRS. NM acknowledges the support provided through Grants-in-Aid for Scientific Research (No. 23681033) from MEXT, Japan, the Nakajima Foundation, and JSPS and F.R.S.-FNRS under the Japan-Belgium Research Cooperative Program.

\expandafter\ifx\csname url\endcsname\relax\def\url#1{\texttt{#1}}\fi
\expandafter\ifx\csname urlprefix\endcsname\relax\def\urlprefix{URL }\fi
\expandafter\ifx\csname href\endcsname\relax\def\href#1#2{#2} \def\path#1{#1}\fi

\bibliographystyle{unsrt}
\bibliography{RochaMasuda}

\clearpage

\begin{center}

\Large Supplementary Material: \\ Random walk centrality for temporal networks \\
\vspace{0.5 cm}
\large Luis E. C. Rocha\dag~and Naoki Masuda\ddag \\
\vspace{0.5 cm}
\small \dag~Department of Public Health Sciences, Karolinska Institutet, Solna, Sweden \\
\small \dag~Department of Mathematics, Universit\'e de Namur, Namur, Belgium \\
\small \ddag~Department of Mathematical Informatics, The University of Tokyo, Tokyo, Japan \\
\small \ddag~CREST, JST, Saitama, Japan \\
\small \texttt{masuda@mist.i.u-tokyo.ac.jp} \\
\vspace{0.2 cm}
\normalsize{\today}

\end{center}

\section{Introduction}

This Supplementary Material contains the results for the in-strength approximation and for the right-moment hypothesis for the SPH, SEX and EMA data sets described in the main text. These results agree with the theoretical predictions described in the main text.

\section{In-strength approximation}\label{sub:in-strength approximation}

\begin{figure*}[ht]
\centering
\includegraphics[scale=0.8]{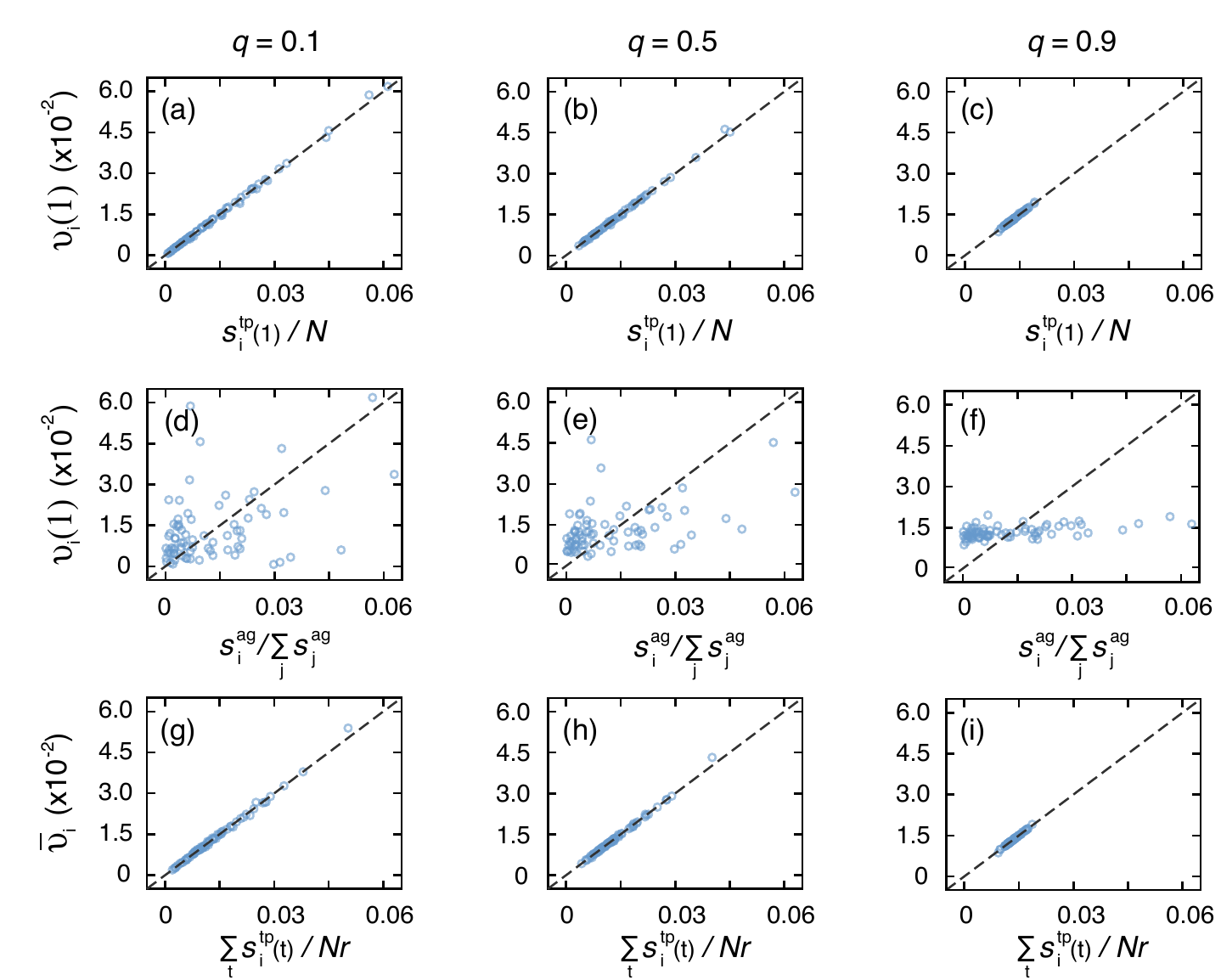}
\caption{\textbf{TempoRank for SPH network.} (a)--(c) Relationship between the stationary density of the temporal transition matrix and the in-strength approximation. Each circle represents a node, and the dashed lines represent the diagonal. (d)--(f) Relationship between the stationary density of the temporal transition matrix and the in-strength of the aggregate network. (g)--(i) Relationship between the TempoRank and the time-averaged in-strength of the effective network. We set $q=0.1$ in (a), (d), (g), $q=0.5$ in (b), (e), (h), and $q=0.9$ in (c), (f), (i). The resolution is $T_{\rm w}= 5$ min.}
\label{fig1S}
\end{figure*}

\begin{figure*}[ht]
\centering
\includegraphics[scale=0.8]{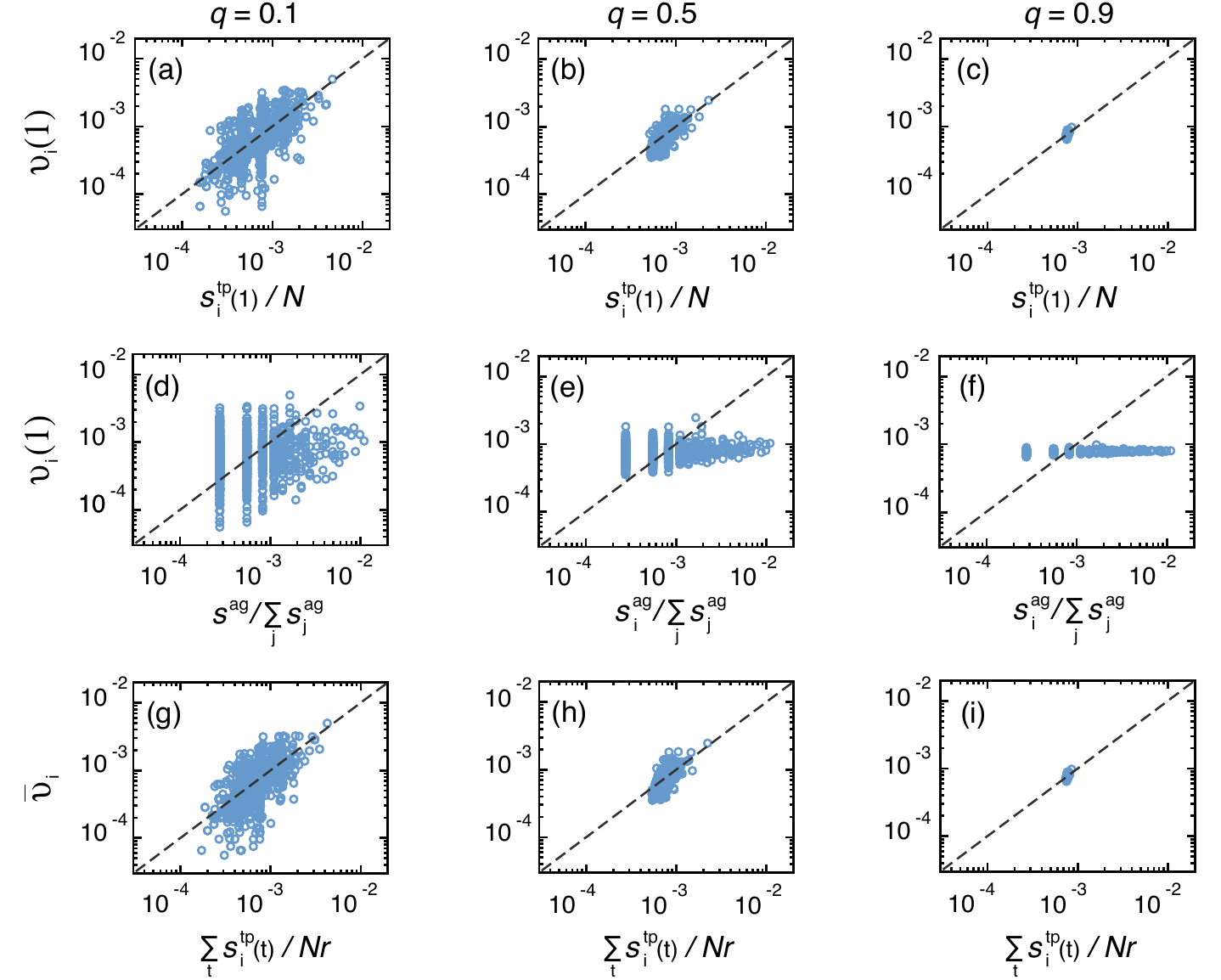}
\caption{\textbf{TempoRank for SEX network.} (a)--(c) Relationship between the stationary density of the temporal transition matrix and the in-strength approximation. (d)--(f) Relationship between the stationary density of the temporal transition matrix and the in-strength of the aggregate network. (g)--(i) Relationship between the TempoRank and the time-averaged in-strength of the effective network. We set $q=0.1$ in (a), (d), (g), $q=0.5$ in (b), (e), (h), and $q=0.9$ in (c), (f), (i). The resolution is $T_{\rm w}= 2$ days. The axes are in log-scale.}
\label{fig2S}
\end{figure*}

\begin{figure*}[ht]
\centering
\includegraphics[scale=0.8]{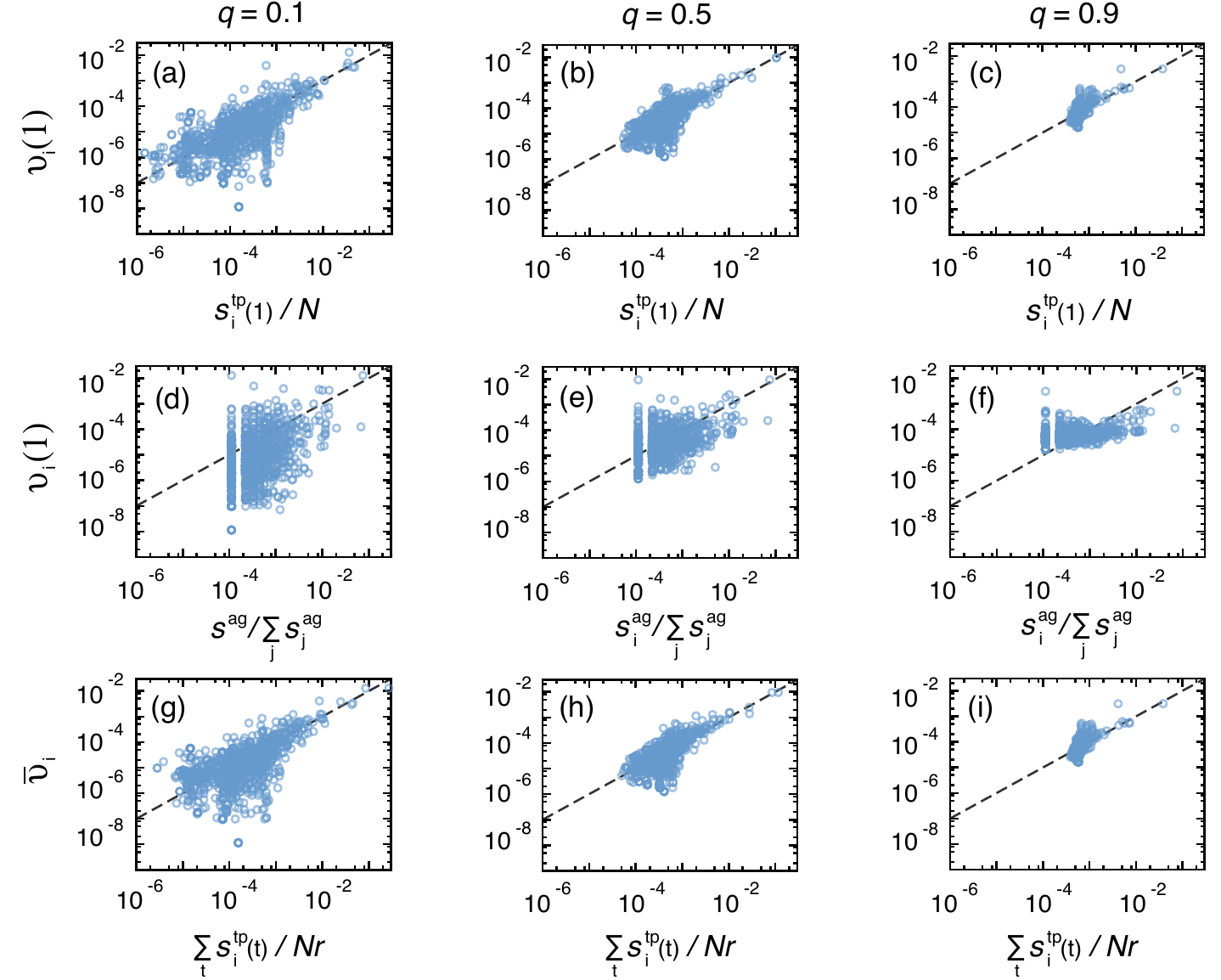}
\caption{\textbf{TempoRank for EMA network.} (a)--(c) Relationship between the stationary density of the temporal transition matrix and the in-strength approximation. (d)--(f) Relationship between the stationary density of the temporal transition matrix and the in-strength of the aggregate network. (g)--(i) Relationship between the TempoRank and the time-averaged in-strength of the effective network. We set $q=0.1$ in (a), (d), (g), $q=0.5$ in (b), (e), (h), and $q=0.9$ in (c), (f), (i). The resolution is $T_{\rm w}= 1$ hour. The axes are in log-scale.}
\label{fig3S}
\end{figure*}

The performance of the in-strength approximator for three values of the sojourn probability $q$ is shown in Fig.~\ref{fig1S} (SPH), Fig.~\ref{fig2S} (SEX) and Fig.~\ref{fig3S} (EMA). The in-strength approximation (Panels (a)--(c)) is accurate for all tested values of $q$ (i.e.\ from 0.1 to 0.9). As is the case for the other data sets, the in-strength of the aggregate network, i.e.\ $s_i^{\rm ag}$, which gives the exact stationary density of the random walk on the aggregate network, is little correlated with $v_i$ for the same three values of $q$ (Panels (d)--(f)). The in-strength approximator for the TempoRank (see the main text) is also strongly correlated with $\overline{v}_i$ (Panels (g)--(i)). Correlation is stronger for the SPH data set in comparison to SEX and EMA data sets which correspond to considerable sparser networks.

\clearpage

\section{The right-moment hypothesis}

\begin{figure*}[ht]
\centering
\includegraphics[scale=0.7]{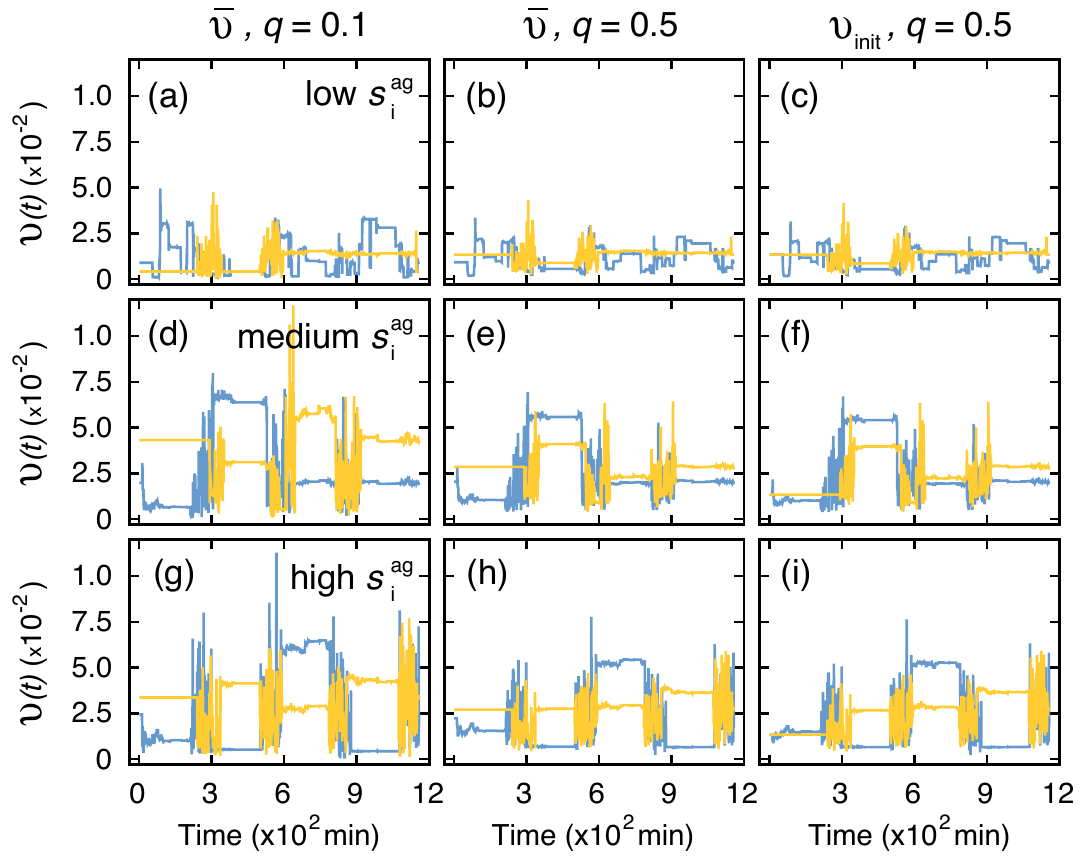}
\caption{\textbf{Time dependence of the stationary density of the random walk for the SPH data set.} In (a), (b), (d), (e), (g), and (h), the density of walkers $\bm v(t)=\overline{\bm v} B(1)\cdots B(t-1)$ is shown. In (c), (f), and (i), the density of walkers in a snapshot $t$ calculated by $\bm v(t) = \bm{v}_{\rm init}(1) B(1)\cdots B(t-1)$, where $\bm{v}_{\rm init}(1) = (1\; \cdots\; 1)/N$, is shown. Each curve corresponds to a node with different $s^{\rm ag}_i$. We set $q=0.1$ in (a), (d), (g), and $q=0.5$ in (b), (c), (e), (f), (h), (i). The resolution is $T_{\rm w}= 5$ min.}
\label{fig4S}
\end{figure*}

\begin{figure*}[ht]
\centering
\includegraphics[scale=0.7]{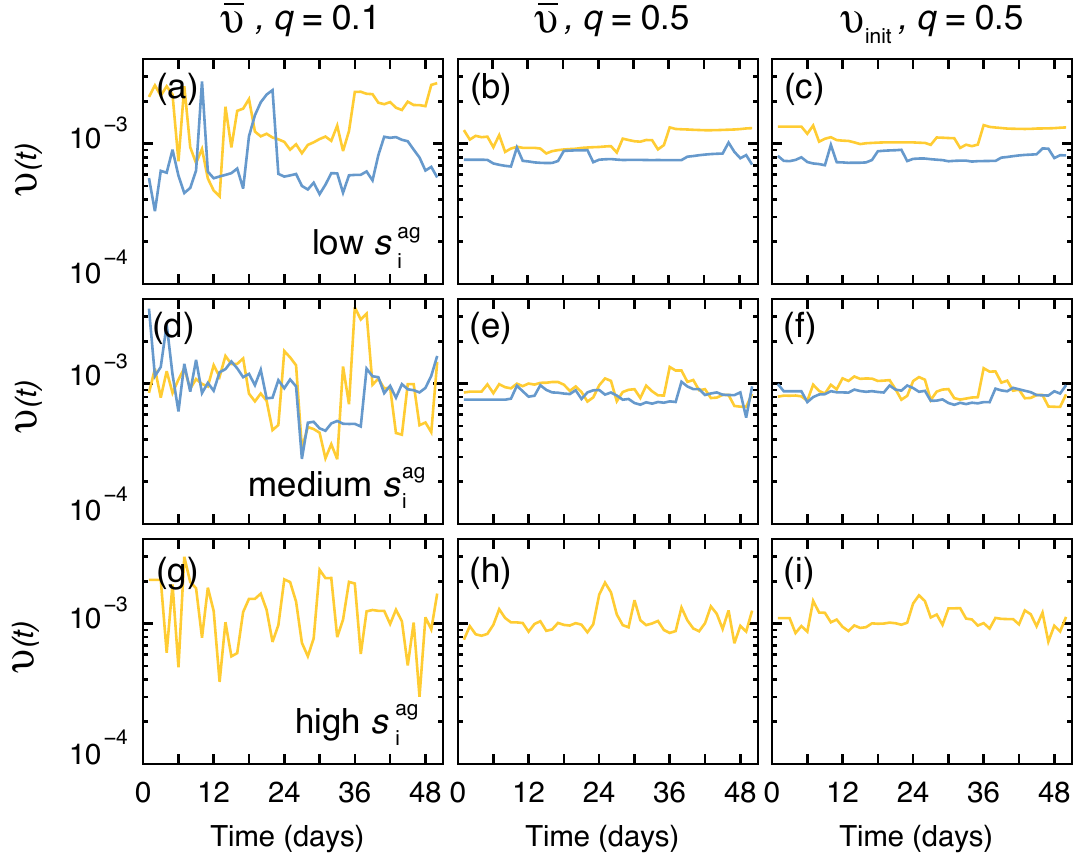}
\caption{\textbf{Time dependence of the stationary density of the random walk for the SEX data set.} In (a), (b), (d), (e), (g), and (h), the density of walkers $\bm v(t)=\overline{\bm v} B(1)\cdots B(t-1)$ is shown. In (c), (f), and (i), the density of walkers in a snapshot $t$ calculated by $\bm v(t) = \bm{v}_{\rm init}(1) B(1)\cdots B(t-1)$, where $\bm{v}_{\rm init}(1) = (1\; \cdots\; 1)/N$, is shown. Each curve corresponds to a node with different $s^{\rm ag}_i$. The resolution is $T_{\rm w}= 2$ days.}
\label{fig5S}
\end{figure*}

\begin{figure*}[ht]
\centering
\includegraphics[scale=0.7]{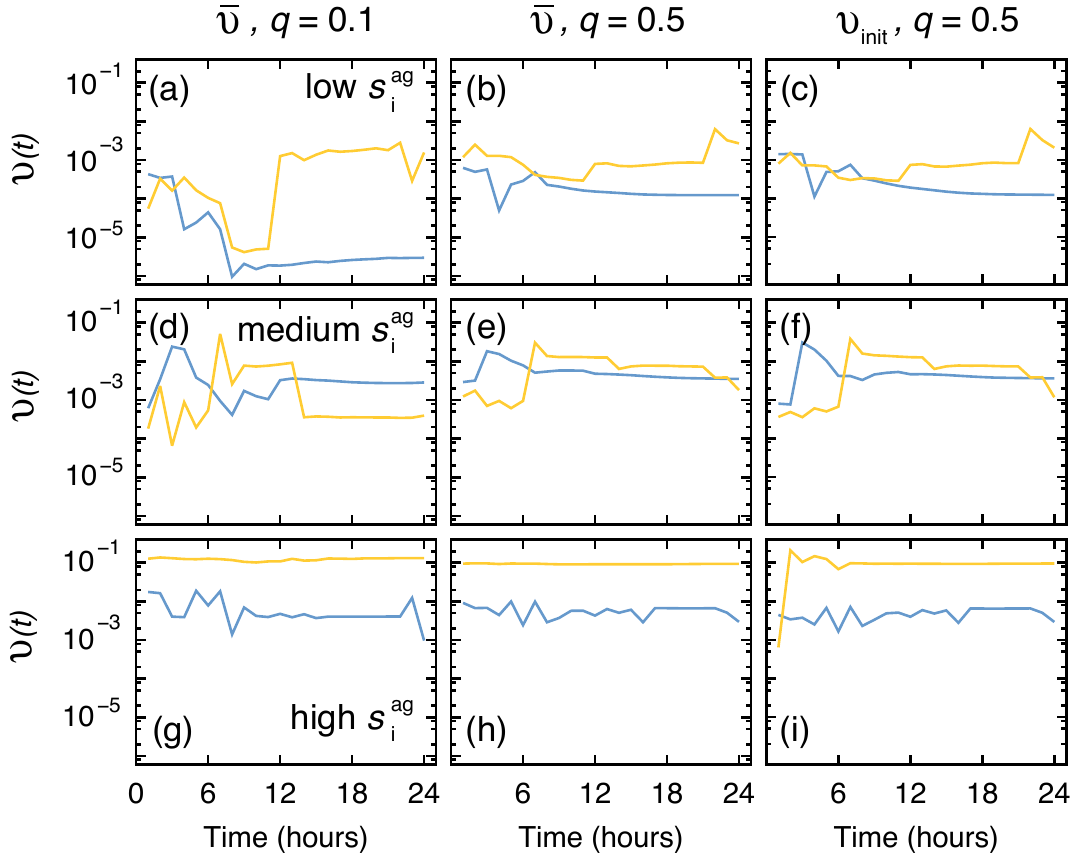}
\caption{\textbf{Time dependence of the stationary density of the random walk for the EMA data set.} In (a), (b), (d), (e), (g), and (h), the density of walkers $\bm v(t)=\overline{\bm v} B(1)\cdots B(t-1)$ is shown. In (c), (f), and (i), the density of walkers in a snapshot $t$ calculated by $\bm v(t) = \bm{v}_{\rm init}(1) B(1)\cdots B(t-1)$, where $\bm{v}_{\rm init}(1) = (1\; \cdots\; 1)/N$, is shown. Each curve corresponds to a node with different $s^{\rm ag}_i$. The resolution is $T_{\rm w}= 1$ hour.}
\label{fig6S}
\end{figure*}

The stationary density at the $t$-th snapshot, i.e.\ $\bm v(t)$, is shown as a function of $t$ in Fig.~\ref{fig4S} (SPH), Fig.~\ref{fig5S} (SEX) and Fig.~\ref{fig6S} (EMA). In each figure, we use two values of $q$ and calculate $v_i(t)$ for representative nodes $i$ with low, intermediate, and high strengths in the aggregate network, i.e.\ the total number of contacts, $s_i^{\rm ag}$. Fluctuations are significant in all cases.

\end{document}